\documentclass{appolb}
\usepackage{epsfig}

\newsavebox{\tabela}
\input epsf.sty

\begin{document}
\title{THERMAL ANALYSIS OF PARTICLE RATIOS AND $p_\perp$ SPECTRA AT  RHIC%
\thanks{Research supported in part by the Polish State Committee for
        Scientific Research, grant 2 P03B 09419}%
}
\author{Wojciech Florkowski, Wojciech Broniowski, Mariusz Michalec
\address{The H. Niewodnicza\'nski Institute of Nuclear Physics, \\
        ul. Radzikowskiego 152,  PL-31342 Krak\'ow, Poland}
}
\maketitle
\begin{abstract}
The thermal model of particle production is used to analyze the
particle ratios and the $p_\perp$ spectra measured recently at
RHIC. Our fit of the particle ratios yields the temperature at the
chemical freeze-out $T_{chem} = 165 \pm 7 $ MeV with the corresponding
baryon chemical potential $\mu^B_{chem} = 41 \pm 5$ MeV.  The quality
of the fit shows that the model works well for RHIC.  The $p_\perp$
spectra are evaluated in an approach which takes into account the
modifications of the initial thermal distributions by the secondary
decays of resonances. All two- and three-body decays are included.
This leads to an effective ``cooling'' of the spectra in the data
region by about 30-40 MeV. We find that the pion spectrum is
characterized by the inverse slope which agrees well with the value
inferred from the RHIC data.
\end{abstract}
  
\section{Introduction}

Recently, a successful description of the particle ratios measured in
relativistic heavy-ion collisions has been achieved in the framework
of so-called thermal models
\cite{pbmags,cest,pbmsps,raf,cr,yg,finland1,finland2,gaz}. An important ingredient of
this approach is the proper inclusion of the hadronic resonances,
whose decays contribute in an essential way to the final
multiplicities of the observed hadrons. In the first part of this
paper the simple thermal model is used to find the optimal values of
the thermodynamic parameters characterizing the ratios measured at
RHIC. In the second part we calculate the $p_{\perp }$ spectra of
hadrons in the thermal model, including all resonances and using an
analytic formula. 
The spectra, besides the original thermal
distribution, receive large contributions from the decaying
resonances. The inclusion of all possible decays is important,
since only in that way one can obtain a clean separation and
identification of other effects such as expansion, flow
\cite{schne,cso,risch,scheibl,tea,huo1,huo2}, or medium
modifications \cite{wfwb,wfwbh,mich}, which influence the final shape of the
$p_{\perp }$ distributions.
The present analysis computes the spectra in a
static fireball, and is introductory to the work of Ref. \cite{wbwf},
where the effects of expansion are incorporated and very good
agreement with experiment is achieved.

\section{Ratios of hadron multiplicities}

In the thermal model the particle densities are calculated 
from the ideal-gas expression
\begin{equation}
n_{i}=\frac{g_{i}}{2\pi ^{2}}\int_{0}^{\infty }\frac{p^{2}\ dp}{\exp \left[
\left( E_{i}-\mu _{chem}^{B}B_{i}-\mu _{chem}^{S}S_{i}-\mu
_{chem}^{I}I_{i}\right) /T_{chem}\right] \pm 1},  \label{ni}
\end{equation}
where $g_{i}$ is the spin degeneracy factor of the $i$th hadron, $
B_{i},S_{i},I_{i}$ are the baryon number, strangeness, and the third
component of isospin, and $E_{i}=\sqrt{p^{2}+m_{i}^{2}}$. The
quantities $ \mu _{chem}^{B},\mu _{chem}^{S}$ and $\mu _{chem}^{I}$
are the chemical potentials enforcing the appropriate conservation
laws. We note that Eq. (\ref{ni}) is used to calculate the
``primordial'' densities of stable hadrons and resonances at the
chemical freeze-out. The final multiplicities receive contributions
from the primordial stable hadrons, as well as from the secondary hadrons
produced by sequential decays of resonances after the freeze-out. We
include {\it all} light-flavor hadrons (with the appropriate branching
ratios) listed in the newest review of particle physics \cite{PDG}.
We neglect the finite-size and excluded volume corrections.%
\footnote{If the sizes of mesons and baryons are equal, the excluded
volume corrections practically cancel in the particle ratios. This is
due to the fact that the overwhelming majority of hadrons is heavy and
may be treated as classical particles.  In this case we may replace
the Fermi-Dirac (Bose-Einstein) distribution function in (\ref {ni})
by the Boltzmann distribution function, for which the excluded volume
corrections factorize. We have checked that the use of the classical
statistics changes the values of the thermodynamic parameters by less
than 3\%.}
The temperature, $T_{chem}$, and the baryonic chemical potential, $\mu
_{chem}^{B}$, are fitted by minimizing the expression $\chi
^{2}=\sum_{k=1}^{n}\left( R_{k}^{\,exp}-R_{k}^{\,therm}\right) ^{2}/\sigma
_{k}^{2}$ , where $R_{k}^{\,exp}$ is the $k$th measured ratio, $\sigma _{k}$
is the corresponding error, and $R_{k}^{\,therm}$ is the same ratio as
determined from the thermal model. The potentials $\mu
_{chem}^{S}$ and $\mu _{chem}^{I}$ are
determined by the two requirements: the initial strangeness of the system is
zero, and the ratio of the baryon number to the electric charge is the same
as in the colliding nuclei.

\savebox{\tabela}{\vbox{
\tabcolsep=1.8mm
\begin{tabular}{|r|r|c|}
\hline
& Thermal Model & Experiment \\ \hline \hline
$T_{chem}$ [MeV] & 165$\pm 7$ &  \\ \hline
$\mu _{chem}^{B}$ [MeV] & \ 41$\pm 5$ &  \\ \hline
$\mu _{chem}^{S}$ [MeV] & \ \ \ \ \ 9 &  \\ \hline
$\mu _{chem}^{I}$ [MeV] & \ \ \ \ \ -1 &  \\ \hline
$\chi ^{2}/n$ & 0.97 &  \\ \hline \hline
$\pi ^{-}/\pi ^{+}$ & $1.02$ & 
\begin{tabular}{ll}
$1.00\pm 0.02$ \cite{phobos}, & $0.99\pm 0.02$\cite{bearden}
\end{tabular}
\\ \hline 
$\overline{p}/\pi ^{-}$ & $0.09$ & $0.08\pm 0.01$ \cite{harris} \\ \hline
$K^{-}/K^{+}$ & $0.92$ & 
\begin{tabular}{ll}
$0.88\pm 0.05$ \cite{caines}, & $0.78\pm 0.12$ \cite{ohnishi} \\ 
$0.91\pm 0.09$ \cite{phobos}, & $0.92\pm 0.06$ \cite{bearden}
\end{tabular}
\\ \hline
$K^{-}/\pi ^{-}$ & $0.16$ & $0.15\pm 0.02$ \cite{caines} \\ \hline
$K_{0}^{\ast }/h^{-}$ & $0.046$ & $0.060\pm 0.012$ \cite{caines,zxu} \\ 
\hline
$\overline{K_{0}^{\ast }}/h^{-}$ & $0.041$ & $0.058\pm 0.012$ \cite
{caines,zxu} \\ \hline
$\overline{p}/p$ & $0.65$ & 
\begin{tabular}{ll}
$0.61\pm 0.07$ \cite{harris}, & $0.54\pm 0.08$ \cite{ohnishi} \\ 
$0.60\pm 0.07$ \cite{phobos}, & $0.61\pm 0.06$ \cite{bearden}
\end{tabular}
\\ \hline
$\overline{\Lambda }/\Lambda $ & $0.69$ & $0.73\pm 0.03$ \cite{caines} \\ 
\hline
$\overline{\Xi }/\Xi $ & $0.76$ & $0.82\pm 0.08$ \cite{caines} \\ \hline
\end{tabular}}}
\begin{table}[t]
\vspace{1.0cm}
\usebox{\tabela}
\caption{Our fit to the particle ratios measured at RHIC.}
\label{tab:results}
\end{table}

Table 1 presents our fit to the particle ratios measured at RHIC. In
our calculation, the identical ratios measured by different groups are
treated separately in the definition of $\chi^{2}$ (number of points
$n=16$). In this way the measurements done by different groups enter
independently. Very similar results are obtained if we first average
the results of different groups to obtain the most likely value for
each considered ratio. Our optimal value of $T_{chem}=165\pm 7$ MeV is
consistent with the value of the critical temperature as inferred from
the lattice simulations of QCD (with three massless flavors
$T_{C}=154\pm 8$ MeV, whereas with two massless flavors $T_{C}=173\pm
8$ MeV \cite{Karsch}). We have also calculated other characteristics of
the freeze-out.  In particular, we find the energy density
$\varepsilon =0.5$ GeV/fm$^{3}$, the pressure $P=$ 0.08 GeV/fm$^{3}$,
and the baryon density $\rho _{B}=$ 0.02 fm$^{-3}$. Our calculation
confirms the Cleymans-Redlich conjecture \cite{cr} that the energy per
hadron at the chemical freeze-out is 1 GeV (our approach yields
$\left\langle E\right\rangle /\left\langle N\right\rangle =$ 1.0
GeV). We observe, however, that the average energy per baryon is much
larger than the average energy per meson: $\left\langle
E_{B}\right\rangle /\left\langle N_{B}\right\rangle =$ 1.6 GeV and $
\left\langle E_{M}\right\rangle /\left\langle N_{M}\right\rangle =$
0.9 GeV.  Similar differences occur also for other heavy-ion
collisions studied at AGS and SPS. They are caused by the different
growth rates of the meson and baryon mass spectra \cite{hag}. In
addition, we find that the ratios $\overline{\Lambda } /\Lambda $ and
$\overline{\Xi }/\Xi $ are practically unaffected by the weak
decays. In conclusion, the thermal model works well for the RHIC
particle ratios, with thermal parameters assuming anticipated values.

\begin{figure}[t]
\epsfysize=7cm
\par
\begin{center}
\mbox{\epsfbox{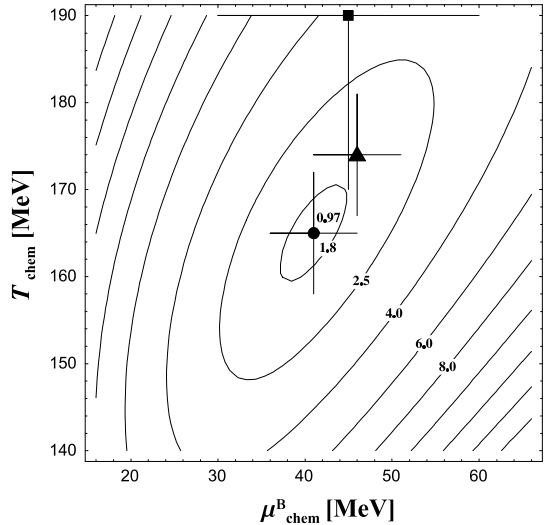}}
\end{center}
\caption{ The contour plot of our $\chi^2/n$ treated as a function of 
$T_{chem}$ and $\mu^B_{chem}$. Our result (black circle),
the result of Ref. \cite{pbmrhic} (black triangle), and the fit of
Ref. \cite{nxu} (black square) are all shown with the corresponding errors.}
\label{figchi2}
\end{figure}

We note that our $T_{chem}$ is 9 MeV lower than 174 MeV of
Ref. \cite{pbmrhic}, and 25 MeV lower than 190 MeV obtained in
Ref. \cite{nxu}. Nevertheless, the results of the three calculations
are consistent within errors, as displayed in Fig.~\ref{figchi2}.  An
interesting feature shown in Fig.~\ref{figchi2} is the characteristic
valley of the optimal parameters. The shape of this valley indicates
that the quality of the fit does not change if we moderately increase
or decrease both $T_{chem}$ and $\mu^B_{chem}$.

\section{Transverse-momentum spectra}

Next, we come to the discussion of the $p_{T}$ spectra. Similarly to
the particle multiplicities, the momentum spectra of the observed
hadrons contain two contributions. The first, ''primordial''
contribution is purely thermal and is described by distribution
functions characterized by the thermodynamic parameters at
freeze-out. The second contribution comes from sequential decays of
the resonances. Since a substantial part of the produced particles
comes from the decays, one may expect that the measured spectra are
significantly changed by this effect. Indeed, it has been already
known for a long time that the resonance decays modify the low-$p_{T}$
spectrum, due to the limited phase space of the emitted particles
\cite{SKH,BSW,weinhold}. So far, however, a rather limited number of
the resonances has been included in such analyses. We take into
account all two- and three-body hadronic decays. In this study we
neglect other effects which can change the spectra, such as expansion
or flow.

Below we sketch our method of dealing with two- and three-body
decays. The initial momentum distribution function of a resonance in
the fireball rest frame, $f(k)$, as well as the momentum distribution
of the emitted particle in the resonance's rest frame, are
isotropic. The latter is true, since we average over all possible
polarization states. Thus, in the case of a two-body decay, the
spectrum of the emitted particle (denoted by index 1) is obtained from
the expression
\begin{equation}
\widetilde{f}_1 \left( \left| {\bf q}\right| \right) = 
b \; \frac{2 J_R +1}{2 J_1+1}
\int d^{3}k\ f\left(
k\right) \int \frac{d^{3}p}{4\pi p^{\ast 2}}\delta \left( \left| {\bf p}%
\right| -p^{\ast }\right) \delta ^{(3)}\left( \hat{L}_{k}{\bf p-q}%
\right) ,  \label{ftilde}
\end{equation}
where ${\bf p}$ is the momentum of the emitted particle in the rest
frame of the resonance, and $\hat{L}_{k}{\bf \ }$is the Lorentz
transformation to the fireball rest frame, 
\begin{equation}
\hat{L}_{k}{\bf p=p}+
[(\gamma _{k}-1) \ v_{k}^{2}\ {\bf v}_{k}\cdot {\bf
p} + \gamma _{k}\ E^{\ast }]{\bf v}_{k}. 
\end{equation}
Here ${\bf k}$ is
the momentum of the resonance in the fireball, ${\bf v}_{k}={\bf k}
/\sqrt{k^{2}+m_{R}^{2}}$, and $\gamma _{k}=\left(
1-v_{k}^{2}\right) ^{-1/2}$. In Eq. (\ref{ftilde}) $p^{\ast }$ is the
magnitude of ${\bf p}$ and $E^{\ast}=\sqrt{p_\ast^2+m_1^2}$. The
standard formula gives 
\begin{equation}
p^{\ast}={
\left[(m_{R}^{2}-(m_{1}-m_{2})^{2})(m_{R}^{2}-( m_{1}+m_{2})^{2})\right]^{1/2} 
\over 2m_{R}},
\end{equation}
where $m_{R}$ is the mass of the resonance, whereas $m_{1}$ and
$m_{2}$ are the masses of the emitted particles.  The quantity $b$ is
the branching ratio for the considered channel, and $J_R$ and $J_1$
are the spins of the resonance and particle 1, respectively.  The
physical interpretation of Eq. (\ref{ftilde}) is clear: the isotropic
distribution of particle 1 in the resonance rest frame, $\delta \left(
\left| {\bf p} \right| -p^{\ast }\right)/(4\pi p^{\ast 2})$, is
boosted to the fireball frame, and there folded with the resonance
distribution $f(k)$.

In order to write Eq. (\ref{ftilde}) in a more compact form, we
do the change of variables:
\begin{equation}
{\bf p}^\prime =  \hat{L}_{k}{\bf p}, \hspace{1.25cm}
d^3p = {E_p \over E_{p^\prime}} \, d^3 p^\prime ={E^\ast \over E_q } d^3 p^\prime.
\label{change}
\end{equation}
The integration over $p^\prime$ in (\ref{ftilde}) becomes trivial and we find 
\begin{equation}
\widetilde{f}_1 \left( \left| {\bf q}\right| \right) = 
b \; \frac{2 J_R +1}{2 J_1+1}  \frac{1}{4\pi p^{\ast 2}} {E^\ast \over E_q}
\int d^{3}k\ f\left(k\right)
\delta \left( \left| \hat{L}_{k}^{-1} {\bf q} \right|-p^{\ast }\right).
\label{ftilde1}
\end{equation}
Since the quantity $ \hat{L}_{k}^{-1} {\bf q}$ is the momentum of
the emitted particle in the reference frame connected with the resonance,
we may rewrite Eq. (\ref{ftilde1}) in the explicitly Lorentz-covariant way
\begin{equation}
\widetilde{f}_1 \left( \left| {\bf q}\right| \right) = 
b \; \frac{2 J_R +1}{2 J_1+1}  \frac{1}{4\pi p^{\ast}} 
{1 \over E_q}
\int d^{3}k\ f\left(k\right)
\delta \left( {k^\mu \cdot q_\mu \over m_R}-E^\ast \right).
\label{ftilde2}
\end{equation}
Here $k^\mu$ and $q^\mu$ are the four-momenta of the resonance and of
the emitted particle, respectively.  We note that Eq. (\ref{ftilde2})
was used previously by \mbox{Sollfrank}, Koch, and Heinz \cite{SKH}. In their
approach, however, no assumptions about the isotropic emission were
made, so $f(k)$ and consequently $\widetilde{f}_1 (q)$ were treated as the
functions of two arguments: rapidity and transverse momentum.
In our approach $f(k)$ and $\widetilde{f}_1 (q)$ depend only on the
magnitude of the three-momenta, and  integration over the polar
and azimuthal angles in (\ref{ftilde2}) can be done analytically.
This leads to a simple expression
\begin{equation}
\widetilde{f}_1\left( q\right) = b \; \frac{2 J_R +1}{2 J_1+1}
\frac{m_R}{2 E_q p^\ast q} \int_{k_{-}(q)}^{k_{+}(q)} dk\, k \,
f\left( k\right),  \label{ftildecom}
\end{equation}
where the limits of the integration are $k_{\pm }(q)=m_{R} \left|
E^\ast q \pm p^\ast E_q \right| / m_1^2 $.  Eq. (\ref{ftildecom}) is a
relativistic generalization of the formula derived in
Ref. \cite{weinhold}. Its simplicity turns out to be especially
important in the numerical treatment of the sequential decays.

In the case of three-body decays we can follow the same steps as above, with
the extra modification connected with the fact that different values of $
p^{\ast }$ are possible now. This introduces an additional integration in
Eq. (\ref{ftilde}). The distribution of the allowed values of $p^{\ast}$ may
be obtained from the phase-space integral 
\begin{equation}
N\int \frac{d^{3}p_{1}}{E_{p1}}\frac{%
d^{3}p_{2}}{E_{p_{2}}}\frac{d^{3}p_{3}}{E_{p_{3}}}\delta \left(
m_{R}\!-\!E_{p_{1}}\!-\!E_{p_{2}}\!-\!E_{p_{3}}\right) \delta ^{(3)}\left( 
{\bf p}_{1}\!+\!{\bf p}_{2}\!+\!{\bf p}_{3}\right) \left| M\right| ^{2}, 
\label{gp1}
\end{equation}
where ${\bf p}_{1},{\bf p}_{2}$ and ${\bf p}_{3}$ are the momenta of
the emitted particles, $E_{p_{1}},E_{p_{2}}$ and $E_{p_{3}}$ are the
corresponding energies (all measured in the resonance rest frame), $M$
is the matrix element describing the three-body decay, and $N$ is the
normalization constant.  For sake of simplicity we assume, similarly
as in \cite{SKH}, that $M$ can be approximated by a
constant. Operationally, the final expression for three-body decays is
a folding of two-body decays over $p^{\ast }$ with a weight following
from elementary considerations based on Eq. (\ref{gp1}).

\begin{figure}[ht]
\epsfysize=8.5cm
\par
\begin{center}
\mbox{\epsfbox{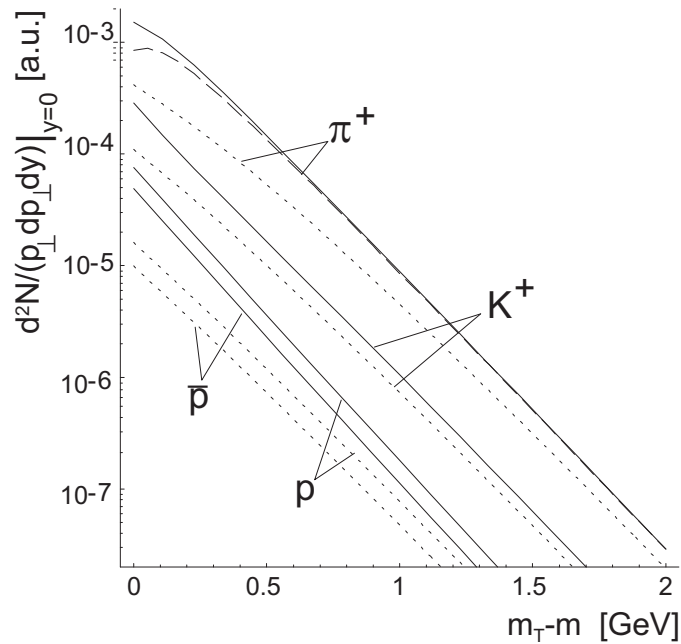}}
\end{center}
\caption{The transverse-mass distributions of pions, kaons, protons and 
antiprotons. }
\label{figspec}
\end{figure}

Our numerical procedure is as follows: We initialize the spectra of
all hadrons as given by the thermal model. Next, we start from the
heaviest particle, and proceed with the decay, thus feeding the
spectra of the products. We repeat this step for all particles, going
down with the mass, untill all resonances are taken into account.  Our
results for the $p_\perp$ spectra are shown in Fig. \ref{figspec}. The
initial thermal distributions are represented by the dashed lines,
whereas the final distributions are represented by the solid lines. In
the case of pions the long-dashed line shows the result of including
two-body decays only. We observe that the final distributions of
hadrons are considerably steeper than the original distributions. We
term this phenomenon as {\em ''cooling'' of the spectra by secondary
decays}. For the pions the initial inverse slope, calculated in the
range 0.3 GeV $ < p_\perp < $ 0.9 GeV \cite{velko}, equals 219 MeV.
Hadronic decays cool it by 34 MeV down to 185 MeV. The last value
agrees well with the measurement of the PHENIX Collaboration
\cite{velko}.
\footnote{Contrary to a naive expectation, the inverse slopes of the
thermal distributions in Fig. \ref{figspec} (dotted lines) do not
correspond to the physical temperature characterizing the thermal
distribution $f$.  This effect is induced by the presence of $m_\perp$ as
a prefactor multiplying the thermal distribution in the expression for
the yield: $dN/(m_T dm_T dy) \sim m_\perp \hbox{cosh}(y) f(m_\perp
\hbox{cosh}(y))$. Since the values of $m_\perp$ are not asymptotic,
the inverse slopes obtained in the region of $p_\perp \sim 0.5 - 1$
GeV are considerably higher than the temperature.} The inverse slopes
obtained for other hadrons are smaller than those inferred from the
data. Definitely, other processes influence the observed spectra.

Our calculation does not show any peaked low-$p_\perp$ enhancement in
the pion spectrum. It would, if for instance we had included only the
$\Delta(1232)$ decays. However, with all decays included, the increase
of the spectrum is uniform due to the fact that different decays
populate different momenta. The shape of the pion spectrum in
Fig. \ref{figspec} is concave, thus the thermal model does not
reproduce the experimental convex shape \cite{velko}.  One can see
that the contributions from the three-body decays to the pion spectrum
are important at small momenta, however, they are smaller than the
contributions from the two-body decays. The impact of three-body
decays on other spectra is negligible.

Certainly, the expansion or flow effects further modify the spectra.
Their inclusion is necessary to obtain the complete agreement with the
data \cite{wbwf}.  However, the effect of ''cooling'' from secondary
decays described in this paper is an important ingredient of any
analysis of momentum spectra in relativistic heavy-ion collisions.


\begin{thebibliography}{99}

\bibitem{pbmags} P. Braun-Munzinger, J. Stachel, J. P. Wessels, and N. Xu, Phys.
Lett. {\bf B344}, 43 (1995); Phys. Lett. {\bf B365}, 1 (1996).

\bibitem{cest} J. Cleymans, D. Elliott, H. Satz, and R. L. Thews, Z. Phys. 
{\bf 74}, 319 (1997).

\bibitem{pbmsps} P. Braun-Munzinger, I. Heppe, and J. Stachel, Phys. Lett. 
{\bf B465}, 15 (1999). 

\bibitem{raf} J. Rafelski, J. Letessier, and A. Tounsi, Acta Phys. Pol.
{\bf B28}, 2841 (1997).

\bibitem{cr} J. Cleymans and K. Redlich, Phys. Rev. Lett. {\bf 81}, 5284 (1998).

\bibitem{yg} G. D. Yen and M. I. Gorenstein, Phys. Rev. {\bf C59}, 2788 (1999).

\bibitem{finland1}  F. Becattini, J. Cleymans, A. Keranen, E. Suhonen, and K.
Redlich, Phys. Rev. {\bf C64}, 024901 (2001).

\bibitem{finland2}  F. Becattini, J. Cleymans, A. Keranen, E. Suhonen, and K.
Redlich, hep-ph/0011322.

\bibitem{gaz} M. Ga\'zdzicki, Nucl. Phys. {\bf A681}, 153 (2001).

\bibitem{schne} E. Schnedermann, J. Sollfrank, and U. Heinz,
Phys. Rev.  {\bf C48}, 2462 (1993).

\bibitem{cso} T. Cs\"org\H{o} and B. L\"orstad, Phys. Rev. {\bf C54},
1390 (1996).

\bibitem{risch} D. H. Rischke and M. Gyulassy, Nucl. Phys. {\bf A697},
701 (1996); Nucl. Phys. {\bf A608}, 479 (1996).

\bibitem{scheibl} R. Scheibl and U. Heinz, Phys. Rev. {\bf C59}, 1585 (1999).

\bibitem{tea} D. Teaney. J. Lauret, and E. V. Shuryak, Phys. Rev. Lett.
{\bf 86}, 4783 (2001).

\bibitem{huo1} P. Huovinen, P. F. Kolb, U. Heinz, P. V. Ruuskanen, and
S. A. Voloshin, Phys. Lett. {\bf B503}, 58 (2001).

\bibitem{huo2} P. Huovinen, nucl-th/0108033.

\bibitem{wfwb}  W. Florkowski and W. Broniowski, Phys. Lett. {\bf B477}, 73
(2000).

\bibitem{wfwbh} W. Florkowski and W. Broniowski, Proceedings of the
International Workshop XXVIII on Gross Properties of Nuclei and Nuclear
Excitations, Hirschegg, Austria, 2000, p. 275.

\bibitem{mich} M. Michalec, W. Florkowski and W. Broniowski, 
Phys. Lett. {\bf 520}, 213 (2001).

\bibitem{wbwf} W. Broniowski and W. Florkowski, Phys.  Rev. Lett. in
print, nucl-th/0106050.

\bibitem{PDG}  Particle Data Group, Eur. Phys. J. {\bf C15}, 1 (2000).

\bibitem{Karsch} F. Karsch, Proceedings of QM2001, Nucl. Phys. {\bf A}
in print, hep-ph/0103314.

\bibitem{hag} W. Broniowski and W. Florkowski, Phys. Lett. {\bf B490},
223 (2000).

\bibitem{phobos} B. B. Back, PHOBOS Collaboration,
Phys. Rev. Lett. {\bf 87}, 102301 (2001).

\bibitem{bearden} I. G. Bearden, BRAHMS Collaboration, Proceedings of
QM2001, Nucl.  Phys. {\bf A} in print.

\bibitem{harris} J. Harris, STAR Collaboration, Proceedings of QM2001,
Nucl. Phys. {\bf A} in print.

\bibitem{caines} H. Caines, STAR Collaboration, Proceedings of QM2001,
Nucl. Phys. {\bf A} in print.

\bibitem{ohnishi} H. Ohnishi, PHENIX Collaboration, Proceedings of
QM2001, Nucl. Phys.  {\bf A} in print.

\bibitem{zxu} Z. Xu, Proceedings of QM2001, Nucl. Phys. {\bf A} in
print, nucl-ex/0104001.

\bibitem{pbmrhic} P. Braun-Munzinger, D. Magestro, K. Redlich, and
J. Stachel, Phys. Lett. {\bf B518}, 41 (2001).

\bibitem{nxu} N. Xu and M. Kaneta, Proceedings of QM2001,
Nucl. Phys. A in print, nucl-ex/0104021.

\bibitem{SKH} J. Sollfrank, P. Koch, and U. Heinz, Phys. Lett. {\bf
252}, 256 (1990).

\bibitem{BSW} G. E. Brown, J. Stachel, and G. M. Welke,
Phys. Lett. {\bf B253}, 19 (1991).

\bibitem{weinhold} W. Weinhold, {\it Zur Thermodynamik des $\pi
N$-Systems}, Diplomarbeit, GSI, Sept. 1995.

\bibitem{velko} J. Velkovska, PHENIX Collaboration, Proceedings of
QM2001, Nucl. Phys. {\bf A} in print, nucl-ex/0105012.


\end{thebibliography}
\end{document}